\documentclass[conference]{IEEEtranSPMB}

\IEEEoverridecommandlockouts
\ifCLASSINFOpdf
\else
\fi
\usepackage{amsmath,graphicx,soul,subcaption,balance}
\usepackage{balance}
\usepackage[numbers,sort&compress]{natbib}

\usepackage{parskip}
\usepackage{float} 

\usepackage{mathptmx}

\usepackage{multirow} 

\usepackage{color}
\usepackage{soul,mathrsfs}
\usepackage[utf8]{inputenc}
\usepackage[T1]{fontenc}
\usepackage{array}
\usepackage[hyphens, spaces, obeyspaces]{url}

\usepackage[papersize={8.5in,11in}, left=1in, right=1in, top=1in, bottom=1in]{geometry}
\newcolumntype{L}[1]{>{\raggedright\let\newline\\\arraybackslash\hspace{0pt}}m{#1}}
\newcolumntype{C}[1]{>{\centering\let\newline\\\arraybackslash\hspace{0pt}}m{#1}}
\newcolumntype{R}[1]{>{\raggedleft\let\newline\\\arraybackslash\hspace{0pt}}m{#1}}

\usepackage[singlelinecheck=false,justification=centering]{caption}
\DeclareCaptionLabelFormat{mylabel}{#1 #2.\hspace{0.7ex}}
\renewcommand{\thetable}{\arabic{table}}

\usepackage{xcolor,colortbl}
\definecolor{light-gray}{gray}{0.83}
\newcommand{\GREY}{\cellcolor{light-gray}\bf} 

\newcommand{\spmbtitlefont}{\fontsize{11.0pt}{11.00pt}\selectfont\bf\vspace{0.7em}}
\newcommand{\spmbauthorfont}{\fontsize{11.0pt}{11.0pt}\selectfont\vspace{0em}}
\newcommand{\subparagraph}{}
\usepackage[compact]{titlesec}
\titleformat{\section}
   {\center\normalfont\sc}{\thesection.}{0.7ex}{}
\titlespacing{\section}{0pt}{2ex}{1.5ex}
\titlespacing{\subsection}{0pt}{1.5ex}{1.2ex}
\titlespacing{\subsubsection}{0pt}{1ex}{0.9ex}
\makeatletter
\renewcommand*{\@seccntformat}[1]{\csname the#1\endcsname .\hspace{0.7em}}
\makeatother

\usepackage{fancyhdr,lastpage}
\fancypagestyle{firststyle}
{
   \fancyhf{}
   \fancyfoot[L]{\scriptsize 978-1-6654-7029-2/22/\$31.00 \copyright2022 IEEE}
   \fancyfoot[R]{\scriptsize December 3, 2022}
   \fancyfoot[C]{\scriptsize IEEE SPMB 2022}
}
\title{\spmbtitlefont Histopathology DatasetGAN: Synthesizing Large-Resolution Histopathology Datasets
{\vspace{-2.4\baselineskip}
}
}
    \author{\spmbauthorfont\IEEEauthorblockN{
    S.A.~Rizvi\textsuperscript{\it 1}, 
    P.~Cicalese\textsuperscript{\it 1}, 
    S.V.~Seshan\textsuperscript{\it 3}, 
    S.~Sciascia\textsuperscript{\it 4},
    J.U.Becker\textsuperscript{\it 2}  and 
    H.V.~Nguyen\textsuperscript{\it 1}
    }
    \vspace{0.9em}
    \IEEEauthorblockA{\spmbauthorfont 
        1. Department of Electrical and Computer Engineering, University of Houston, Houston, TX, USA \\
        2. Institute of Pathology, University Hospital of Cologne, Cologne, Germany \\
        3. Weill Cornell Medicine, Cornell University, New York City, New York \\
        4. Center of Research of Immunopathology and Rare Diseases and SCDU Nephrology \\
        and Dialysis, Department of Clinical and Biological Sciences, University of Turin, Turin, Italy \\
        \{srizvi10, pcicalese, hienvnguyen\}@uh.edu, svs2002@med.cornell.edu, \\
        savino.sciascia@unito.it, janbecker@gmx.com
    }
}

\newcommand{\PaperTitleSummary}{S.A.\ Rizvi, et al.: Histopathology DatasetGAN ...}

\begin{document}

\IEEEaftertitletext{}
\maketitle

\begin{abstract}

Self-supervised learning (SSL) methods are enabling an increasing number of deep learning models to be trained on image datasets in domains where labels are difficult to obtain. These methods, however, struggle to scale to the high resolution of medical imaging datasets, where they are critical for achieving good generalization on label-scarce medical image datasets. In this work, we propose the Histopathology DatasetGAN (HDGAN) framework, an extension of the DatasetGAN semi-supervised framework for image generation and segmentation that scales well to large-resolution histopathology images. We make several adaptations from the original framework, including updating the generative backbone, selectively extracting latent features from the generator, and switching to memory-mapped arrays. These changes reduce the memory consumption of the framework, improving its applicability to medical imaging domains. We evaluate HDGAN on a thrombotic microangiopathy high-resolution tile dataset, demonstrating strong performance on the high-resolution image-annotation generation task. We hope that this work enables more application of deep learning models to medical datasets, in addition to encouraging more exploration of self-supervised frameworks within the medical imaging domain. 

\end{abstract}


\IEEEpeerreviewmaketitle    
\thispagestyle{firststyle}  
\section{Introduction}
\label{sec:intro}

Deep learning is powering an increasing number of medical image segmentation applications, accelerating the field past previous statistical and machine learning methods \cite{ref1}. Semantic segmentation, which aims to make structural features more apparent in images, is a crucial part of aiding clinicians in making more accurate diagnoses and improving computer-physician interaction. Acquiring the large medical image datasets required for training deep models, however, poses difficult challenges. Medical datasets are 1) expensive to gather, 2) require very time-consuming annotation, often at expert-level, and 3) often come in large file formats, making it computationally challenging to train large models on.

Semi-supervised learning (SSL) allows for learning under these limitations, utilizing more widely available unlabeled data in order to augment datasets and allow the training of larger, more data-hungry models. Current SSL literature includes pseudo-labeling methods \cite{ref2} which attempt to create labels on unlabeled data, and consistency regularization methods which encourage networks to output consistent predictions for augmented data. Contrastive methods \cite{ref3} define an unsupervised loss term on unlabeled examples which allows for the training of feature extractors. These can then be trained on labeled examples, reducing the overall number of labels required for training.

Generative models have been shown to learn a powerful and detailed latent space representation \cite{ref4} while learning to generate images following the original data distribution. \cite{ref4} proposes a semi-supervised framework for pixel-wise classification in images, utilizing a Multilayer Perceptron (MLP) to classify the GAN's latent features into a segmentation map. This framework, however, requires large memory and computational resources which limits its scalability to domains with larger image sizes. We aim to extend this framework and increase its applicability to medical images by reducing the computational requirements needed to learn off of the features learned by the generative models.

In this work, we propose Histopathology DatasetGAN (HDGAN), a framework for generating high-resolution image-annotation pairs for augmenting medical image datasets. We focus on the scalability of the framework to high-resolution images to accommodate the large images sizes typically seen in the field of medical imaging. By utilizing memory-mapped arrays and reducing the number of latent features extracted from the generative model we drastically reduce the memory consumption and computational requirements for the framework, allowing us to generate image-latent pairs at a much higher resolution. We demonstrate the effectiveness of our framework on a Thrombotic Microangiopathy (TMA) large-resolution tile dataset, and expect that this work will open the doors for further exploration and adoption of semi-supervised training strategies for medical image analysis tasks.


\section{Related Works}

Semi-supervised learning aims to train more powerful and data-hungry networks using a mix of labeled and unlabeled data, taking advantage of the large quantities of unlabeled data available in domains where obtaining labels for data is expensive. Pseudo-labeling \cite{ref2, ref6} and consistency regularization \cite{ref3, ref6} have shown success on semantic segmentation tasks, utilizing unlabeled data by first training a network on the small labeled dataset, and then retraining on both real labels and highly confident pseudolabels predicted on the unlabeled data.

Generative networks learn a latent featurespace while learning to synthesize images, resulting in latent features containing rich semantic information which can aid in semi-supervised tasks. Several works have explored using GAN latent features in order to aid in semi-supervised learning tasks for semantic segmentation. \cite{ref8} proposes a framework for projecting the latent feature maps of a generative network onto semantic segmentation maps using a lightweight decoder built off of convolutional and residual blocks. DatasetGAN \cite{ref4} proposed a pixel-level framework that used an ensemble of MLPs to directly classify pixel-level concatenated features from the entire latent space of the generative network into a segmentation mask. Further works building off of DatasetGAN focused on expanding the diversity of the generated dataset \cite{ref9} and augmenting the generator with a label synthesis branch to directly have the generator predict segmentation masks \cite{ref10}.

Semi-supervised semantic segmentation has been explored in many works, due to the high annotative cost and expertise needed for annotation within the medical domain. In \cite{ref11}, the authors predicted pseudo-labels on unlabeled images and set up an optimization problem to alternately update network parameters and predict segmentations on unlabeled medical images. Other works pursued adversarial \cite{ref12} or transformation-consistency training \cite{ref13} in order to utilized unlabeled medical images. In \cite{ref10}, the authors evaluated their image-annotation generative framework on chest X-ray and skin lesion segmentation, showing strong generalization capabilities with their fully generative approach. We focus back on the original DatasetGAN framework, focusing on improving the memory consumption of the framework. \cite{ref14} presented a preliminary analysis of the applicability of DatasetGAN to the medical imaging domain, making three recommendations to the framework in order to make its application to the medical image more tractable. The first two recommendations involved upgrading the backbone generative network to StyleGAN2-ADA to improve training in limited data regimes, and the third recommendation involved reducing the number of latent features extracted from the generative network while maintaining segmentation mask quality. All three recommendations are satisfied in our extension to the DatasetGAN framework, in addition to our use of memory-mapped arrays to make GPU-based training more feasible.


\begin{figure*}[t]
\vspace{0.5em} 
    \centering
    \captionsetup{justification=centering}
    \includegraphics[width=0.95\linewidth,height=8cm]{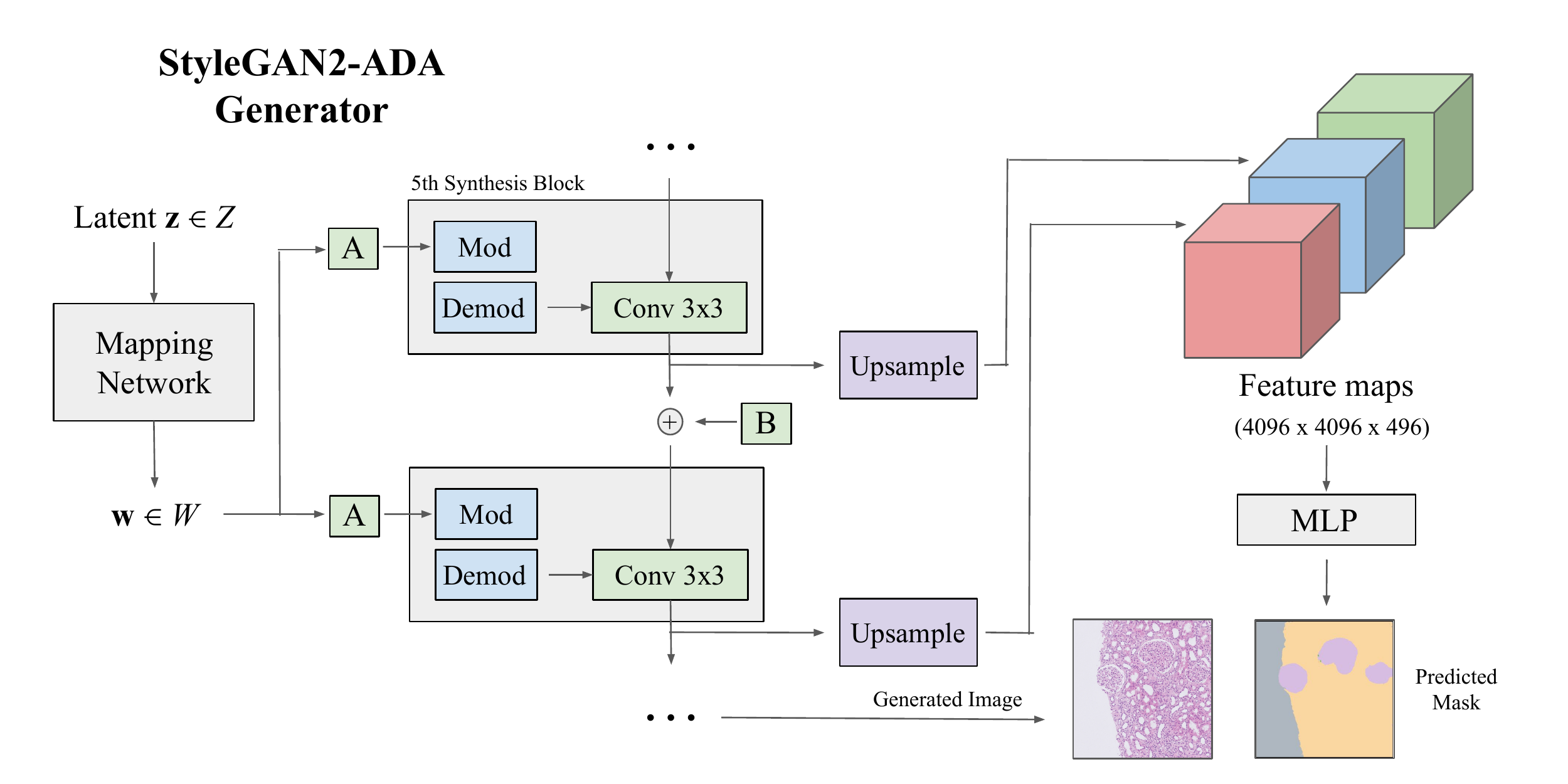}
    \caption{Histopathology DatasetGAN framework overview}
    \vspace{-0.75em} 
    \label{fig:fig1}
\end{figure*}

\section{Data}

The dataset for our experiments consisted of 1,577 whole slide images (WSIs) taken from 100 native kidney biopsies; 50 with a diagnosis of thrombotic microangiopathy (TMA) and 50 mimickers of TMA, with similar or overlapping histopathological features in which a different diagnosis was made. These mimicker diagnoses included necrotising small vessel vasculitis, cryoglobulinemic glomerulopathy, light chain deposition disease, severe benign nephrosclerosis, collapsing type focal and segmental glomerulosclerosis and Bevacizumab-associated obliterative glomerulopathy. The WSIs are taken from the three medical centers (Cologne, Weill-Cornell and Turin) and included a combination of at least three of the four diverse histopathological stainings hematoxylin-eosin (HE), periodic-acid Schiff (PAS), Jones silver and trichrome, all scanned with a x40 objective. In this work, we pre-process the WSIs by first tiling each WSI and extracting any 4096 x 4096 tiles which contained morphological compartments of interest. We then filtered out tiles which consisted of mostly whitespace, resulting in a dataset of 32,732 tiles. We split the tiles into 5 folds and train our StyleGAN2-ADA backbone on 4 of the folds, leaving out one fold as a holdout.

\section{Methodology}

Our method builds off of the original DatasetGAN framework \cite{ref4}, with modifications which make it more suited for high-resolution medical images. The core idea is, given a latent vector z, we can extract the latent features from a generative network when the latent vector is passed through the generator. These feature maps contain semantic information about the image being generated, and therefore are useful for pixel-wise segmentation tasks. The feature maps which are extracted at different resolutions from the generator network are upsampled to match the output image size, yielding a feature map tensor with the same height and width dimensions as the output image. With annotations on a small number of generated images, a smaller classifier can then be trained to predict segmentation maps in a pixel-wise fashion from the extracted feature maps. This provides the framework for generating image-latent pairs.

\begin{figure*}[t]
    \centering
    \includegraphics[width=0.95\linewidth]{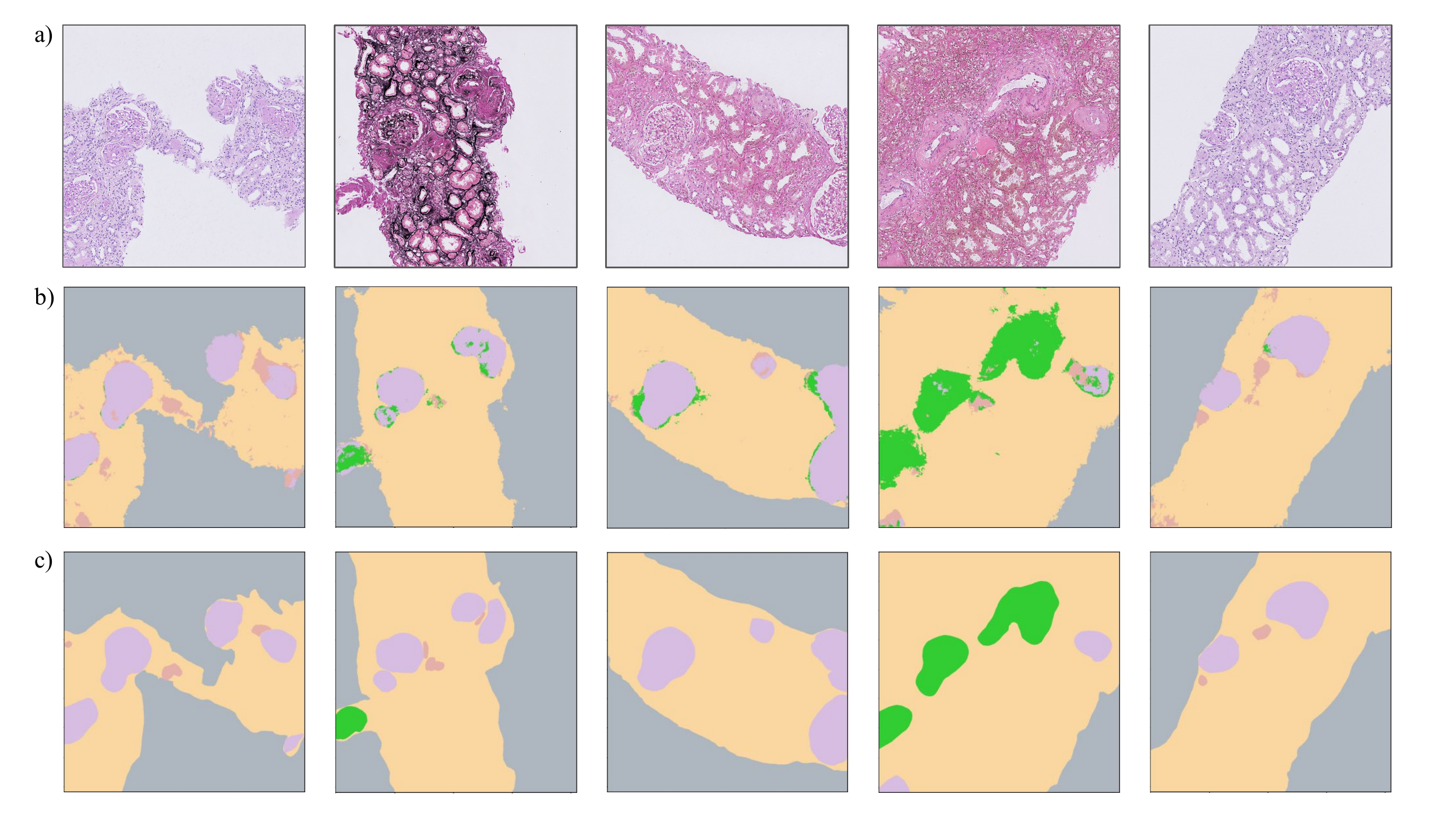}
    \captionsetup{belowskip=6.5pt} 
    \caption{Examples of synthesized images and annotations from the Histopathology DatasetGAN framework. Row a) represents generated TMA 4096x4096 tiles, with row b) and row c) showing the corresponding predicted segmentation map and ground truth segmentation map, respectively.}
    \label{fig:fig2}
\end{figure*}

We extend this framework to work with high-resolution histopathology images by making several adjustments. First, for our generative backbone we use StyleGAN2-ADA \cite{ref5}, which has been proven to work well in limited data settings with histopathology images. Second, unlike \cite{ref4}, we choose to only extract and upsample latent features from the last 5 blocks in the synthesis module of the generator network, as shown in Figure\ \ref{fig:fig1}. This yields feature maps which contain more fine-grained semantic information. These later features are sufficient for our pixel-wise segmentation task, and drastically cut the amount of computation and upsampling required. The computational burden is still too high, however, to fit the entire feature map dataset into memory while training when dealing with large histopathology images. We therefore use memory-mapped arrays during training, which allows for only those portions of the feature map tensor which are currently in use to be loaded from disk and into memory, reducing the memory requirement of the framework. This allows us to scale our framework to 4096 x 4096 TMA tile segmentation, with samples shown in Figure\ \ref{fig:fig2}.

\section{Experimental Design}
We evaluate the HDGAN framework on a 5-class semantic segmentation task on the TMA tile dataset. First, we generate 500 images with a truncation of 0.7 using our trained StyleGAN2-ADA network, saving the corresponding latent vectors. 36 of the 500 images were then selected for our dataset by an expert nephropathologist, who provided pixel-wise annotations of each morphological compartment for these 36 images. We randomly chose 16 images as the training set for the pixel-level classifiers, leaving 4 images for validation and another 16 images as the test set. For baseline experiments, we refer readers to the original DatasetGAN paper \cite{ref4} for comparisons to semi-supervised baselines.

For our classifier, we use a simple 3-layer Multilayer Perceptron (MLP) with ReLU activation, Batch Normalization \cite{ref15}, and Dropout \cite{ref16} following each linear layer. We train the classifier using vanilla stochastic gradient descent with a learning rate of 0.0001. For our loss function, we use CrossEntropy loss.

\section{Results}

In Table\ \ref{table:table1} we report the class-wise pixel-level accuracy of the pixel-classifier on our test set. As expected, the pixel-level classifiers are able to achieve high classification accuracy on several morphological compartments, with an average Dice coefficient of 0.92 on the test set images. The classifiers struggle, however, when morphological structures in generated images resemble two different compartments, often misclassifying arteriole pixels as either glomeruli or artery. This semantic blending was confirmed by our expert annotator, sometimes forcing the annotator to choose between different class labels when a generated compartment had features of both present.

\begin{table}[b!]
%
\centering
\caption{Class-wise Pixel Accuracy}
\label{table:table1}
 \begin{tabular}{|C{0.25\linewidth}|C{0.25\linewidth}|} 
 \hline
{\GREY Class} & {\GREY Accuracy (\%)} \\\hline
Whitespace & 68.42\\\hline
Cortical Tubulointerstitium & 66.57\\\hline
Glomerulus & 82.89\\\hline
Arteriole & 52.09\\\hline
Artery & 77.82\\\hline
\end{tabular}
\end{table}

\section{Summary}
In this paper, we proposed an extension of the DatasetGAN framework with several adaptations making the framework more computationally tractable for large medical images. By utilizing only a subset of latent features from the generative network and using memory-mapped arrays, we greatly reduce the memory requirements of the framework. We also update the backbone of the framework to the StyleGAN2 architecture with adaptive discriminator augmentation, which allows for more stable training in limited data settings as is often encountered in medical imaging domains.



\section{Future Work}

The DatasetGAN framework suffers limitations when the generative backbone is not trained well on the original dataset. This manifested itself in the produced segmentation maps as noisy contours around compartments, where the semantic meaning held in the feature space of the generative networks was not decisively one compartment or another. Future work exploring the latent space of generative networks in relation to semantic meaning of morphological compartments could help disentangle the feature space and result in cleaner segmentation maps. In addition, more works exploring generative frameworks that directly generate the segmentation map such as in \cite{ref10} can help advance the frameworks for dataset augmentation through semi-supervised learning.


\section*{Acknowledgements}

The research in this work was supported by the National Science Foundation and the National Institute of Health. The content represents the views of the authors, not necessarily the views of the funding agencies.


\newpage
\setlength{\bibsep}{2.5pt plus 0ex}
\footnotesize
\bibliographystyle{IEEEbibSPMB}
\bibliography{IEEEabrv,IEEESPMB}

\end{document}